\documentstyle [12pt]{article}
\begin{document}
\baselineskip 24.0pt

$      $
\vspace{0.4in}
\begin{center}
 {\Large\bf Coulomb Correlations and Instability\\
	 of Spinless Fermion Gas in 1D and 2D\\}
   \vspace{0.5truein}
 {\large
   Hua Chen$^{*}$~~~and~~~Daniel Mattis}\\
 {\large Department of Physics\\University of Utah
			      \\Salt Lake City, Utah 84112\\}
\end{center}

\vspace{0.4in}

\begin{abstract}
\large{
We study the stability of the ordinary Landau Fermi liquid phase for
interacting, spinless electrons. We require causality and demand that
the Pauli principle be obeyed. We find a phase diagram determined by
two parameters: the particle density and the interaction strength.
We find that the homogeneous, constant density Fermi liquid phase of
a spinless Fermion gas is {\em never} stable in 1D, but that it may
have a restricted domain of stability in 2D.
}
\end{abstract}

\vspace{2cm}
\rule{5cm}{0.2mm}\\
(*) New address: Department of Computer Science and Applied Physics
Laboratory, The Johns Hopkins University, Baltimore, MD 21218

\newpage
\underline{INTRODUCTION} The theory of the homogeneous electron gas, the
``Fermi liquid'', plays a very important role in study of metallic
properties in real solids. For even within the continuous
{\em jellium model}, Wigner [1] argued long ago that at low density, the
translational symmetry of this phase is broken and electrons are localized
in a lattice to lower the potential energy. Qualitatively speaking, as the
mean interparticle distance $r_s\propto 1/n^{1/d}$ increases, i.e. density
$n$ decreases, the Coulomb potential energy ($\propto 1/r_s$)
of such an electron gas will eventually dominate the kinetic energy
($\propto 1/r_s^2$). As a result, the homogeneous Fermi liquid phase
with a well-defined Fermi surface is unstable in the low density limit. The
nature of the phase which replaces it is, however, a matter of conjecture.
The localized particles in a Wigner lattice are one possibility. A two-phase
regime (high density/low density) is another. The {\em marginal} Fermi
liquid [2] and the quantum Fermi liquid [3] in addition to a variety of
``anyons'' are distinct possibilities in low dimensions. In 3D, there have
been a number of estimates of the critical $r_{s}$, as summarized by
Care and March [4].

Sometime ago Mattis [5] proposed a lattice model which incorporates the
``granular'' nature of electronic states, reflecting the finite size of
the Wannier orbitals of the conduction band of a simple metal. He
found that Coulomb interactions alone could lead to instability
of ordinary paramagnetic Fermi liquid against various interesting
phases, including charge-density waves. Following a theorem of Ginzburg
and Kirzhnits [6], he proposed a loose but general stability criterion,
{\em viz}.: the effective two-body interaction $v(q)$ is required to be
non-negative for all wave vector $q$.

In this paper, we propose a refinement taking into account the short-range
correlation in addition to the causality requirement considered previously.
The application is to {\em spinless} electrons living on a simple
$d$-dimensional cubic structure undergoing mutual Coulomb interactions.
In contrast to the {\em jellium model} where the only length scale is
$r_s$, there appears an additional parameter in the lattice length $a$.
(In the case of {\em real} electrons with spins, the magnetic degrees of
freedom interfere with the charge dynamics to make the analysis much more
complicated.)

The stability criterion of a spinless Fermi liquid will be based on
simultaneously satisfying two requirements: (a) by causality, the inverse
static dielectric function $1/\varepsilon(q,0)<1$; (b) by the Pauli
principle, the zero-range pair correlation function of
spinless Fermions $g(0)=0$. Condition (a), the ``Ginzburg-Kirzhnits
criterion'' [6], is proved using the Kramers-Kronig relation:
\begin{eqnarray}
\mbox{Re}\{\frac{1}{\varepsilon(q,\omega)}\}=1+\frac{1}{\pi}
\int^{\infty}_{0}\mbox{P.}\frac{d\omega'^2}{\omega'^2-\omega^2}
\mbox{Im}\{\frac{1}{\varepsilon(q,\omega')}\}
\end{eqnarray}
where ``P.'' stands for principal value. As
$\mbox{Im}\{\varepsilon(q,\omega)\}\ge 0$, it follows that
the reciprocal static dielectric function $1/\varepsilon(q,0)<1$.

The function $g(r)$ is defined [7] as the probability of finding two
particles separated by a distance $r$, normalized to be $1$ asymptotically
at large $r$. Fortunately $g(r)$ can {\em also} be determined by the
dielectric function according to a well-known many-body identity
(see Eq.(1.6.2), (5.4.12) in reference [7]):
\begin{eqnarray}
g(r)=1+\frac{1}{n}
(\frac{1}{N}\sum_{q}e^{i\vec{q}\cdot\vec{r}}S(q)-1)
\end{eqnarray}
where
\begin{eqnarray}
S(q)=\int^{\infty}_{0}\frac{d\omega}{\pi}\cdot
\frac{1}{nv(q)}\mbox{Im}
\{\frac{-1}{\varepsilon(q,\omega)}\}
\end{eqnarray}
and $v(q)$ is the two-body vertex function.

It is well-known that condition (b) is {\em never} satisfied within
the {\em jellium} model at low density. The calculated [7,8] pair
correlation function $g(r)$ of the continuum theory inevitably turns out
to be negative at small $r$, evidence of an unphysical excess of exchange
and correlation holes around each electron. However, this defect is easily
cured when electrons live on a discrete lattice where the short-range,
on-site interaction and the long-range, off-site Coulomb interaction can be
treated separately. For we have observed that within RPA, condition (b) is
equivalent to the requirement that the ground state energy be stationary
with respect to a fictitious zero-range interaction parametrized by a new
coupling constant $\alpha$. This appears to cure a major flaw in the RPA,
allowing it to be used for semi-quantitative purposes.

\underline{SPINLESS FERMIONS} The extended Hamiltonian is:
\begin{eqnarray}
H=\sum_{k}(\epsilon_k-\mu)c^{\dagger}_{k}c_{k}
 +\frac{1}{2}\sum_{i\neq j}U_{ij}n_in_j
 +\frac{1}{2}\sum_{i}\alpha n_i(n_i-1)
\end{eqnarray}
The third term $\sum_{i}n_i(n_i-1)=0$ vanishes for spinless Fermions
and $\alpha$ is the corresponding undetermined Lagrange multiplier. The
$c's$ are Fermion operators and $n_i=c^{\dagger}_ic_i$ is the particle
number operator on site $i$. In 1D, $\epsilon_k=-2t\cos ka$ ($a$
the lattice length) and $t$ is the transfer integral. The chemical
potential $\mu=\epsilon_{F}$, the Fermi level. The two-body Coulomb
interaction $U_{ij}=e^2/r_{ij}$, $r_{ij}=|i-j|a$; generalization to
higher dimensional hypercubic lattices is straightforward.

To find the ground state energy, we rewrite $H$ in momentum space
keeping careful track of all the terms:
\begin{eqnarray}
H=\sum_{k}(\epsilon_k-\mu)c^{\dagger}_{k}c_{k}
 +\frac{1}{2N}\sum_{q\neq 0}(\alpha+\Delta V(q))\rho(q)\rho(-q)
 -\frac{1}{2}\alpha Nn(1-n)
\end{eqnarray}
where the generalized Ewald sum
\begin{eqnarray}
\Delta V(q)=\sum_{i\neq 0} e^{i\vec{q}\cdot\vec{r_i}}\frac{e^2}{r_{i}}
\end{eqnarray}
and charge density operator $\rho(q)=\sum_k c^{\dagger}_{k+q}c_{k}$.
$N_e$ is the total particle number, $N$ the total number of cells and
$n=N_e/N$ is the particle density. In the usual way, we estimate the
correlation function
\begin{eqnarray}
<\rho(q)\rho(-q)>_{\omega}=\frac{-\Pi(q,\omega)}
{1-(\alpha+\Delta V(q))\Pi(q,\omega)}
\end{eqnarray}
within the RPA approximation by taking $\Pi\approx\Pi^0$, the
Lindhard polarization function [7]. Then the ground state energy is given
by an integral over the coupling constant yielding:
\begin{eqnarray}
E_G=E_0+\frac{1}{2}\mbox{Im}\sum_{q\neq 0}\int^{\infty}_{0}
 \frac{d\omega}{\pi}\log(1-(\alpha+\Delta V(q))\Pi^0(q,\omega))
-\frac{1}{2}\alpha Nn(1-n)
\end{eqnarray}
with $E_0$ the energy of free Fermions. Optimizing $E_G$ with respect
to $\alpha$,
\begin{eqnarray}
\frac{\partial E_G}{\partial\alpha}=0
\end{eqnarray}
we find:
\begin{eqnarray}
\frac{1}{N}\sum_{q\neq 0}S_{\alpha}(q)=1-n
\end{eqnarray}
where
\begin{eqnarray}
S_{\alpha}(q)=\int^{\infty}_{0}\frac{d\omega}{\pi}\cdot
\frac{1}{n(\alpha+\Delta V(q))}\mbox{Im}
\frac{-1}{1-(\alpha+\Delta V(q))\Pi^0(q,\omega)}
\end{eqnarray}
is the structure function. The dielectric function which appears in
Eqs.(1)-(3) is:
\begin{eqnarray}
\varepsilon_{\alpha}(q,\omega)=1-(\alpha+\Delta V(q))\Pi^0(q,\omega),
\hspace{1cm} \mbox{in RPA.}
\end{eqnarray}
By combining Eqs.(2) and (10) we find:
\begin{eqnarray}
g_{\alpha}(0)=1+\frac{1}{n}
(\frac{1}{N}\sum_{q\neq 0}S(q)-1)=0
\end{eqnarray}
Therefore
\begin{eqnarray}
\frac{\partial E_G}{\partial\alpha}=0 \hspace{1cm}\Leftrightarrow
\hspace{1cm} g_{\alpha}(0)=0
\end{eqnarray}
and the stationary condition enforces the Pauli principle. Using the value
of $\alpha$ which satisfies this criterion, we now identify
$V_{\alpha}(q)=\alpha+\Delta V(q)$ as the ``best'' effective two-body
vertex function. Since $\Pi^0(q,0)<0$, Ginzburg and Kirzhnits' criterion
$\varepsilon_{\alpha}(q,0)=1-V_{\alpha}(q)\Pi^0(q,0)=
1+V_{\alpha}(q)|\Pi^0(q,0)|>1$ implies $V_{\alpha}(q)>0$ for all $q$
for the system in the Fermi liquid phase, just as in reference [5]. The
interpretation of our result is that the ``best'' choice of $\alpha$
compensates for the errors in RPA at short distances, and transforms it
into a more reliable procedure.

\underline{NUMERICAL RESULTS} The problem has thus been reduced to
to find a solution for $\alpha$ which satisfies both
\begin{eqnarray}
V_{\alpha}(q)=\alpha+\Delta V(q)>0,
\hspace{1cm}\forall q,
\hspace{1cm}\mbox{and}~~~ g_{\alpha}(0)=0
\end{eqnarray}
We will first solve for $\alpha$, then plug it into the inequality
to check whether it is satisfied. If not, \underline{some} symmetry
breaking must occur.

We present the numerical results in 1D and 2D. In 1D, the generalized Ewald
summation is:
\begin{eqnarray}
\Delta V(q) &=& \sum_{n\neq 0}e^{iq\cdot na}\frac{e^2}{|na|}\nonumber\\
&=& -\frac{e^2}{a}\ln|4\sin^2\frac{qa}{2}|
\end{eqnarray}
and it has a minimum $\Delta V(\pi/a)=-1.386 e^2/a$.
In addition, the retarded polarization function $\Pi^0(q,\omega)$
can also be evaluated analytically, using the tight-binding spectrum
$\epsilon_k=-\cos ka$ (taking $2t=1$ as our unit from now on). By
symmetries $\Pi^0(q,\omega)=\Pi^0(-q,\omega)$,
$\mbox{Re}\Pi^0(q,-\omega)=\mbox{Re}\Pi^0(q,\omega)$,
$\mbox{Im}\Pi^0(q,-\omega)=-\mbox{Im}\Pi^0(q,\omega)$, and because of
periodicity in $q$, we only need to consider $\omega>0$ and $0<qa<\pi$.
For $\omega>2\sin\frac{qa}{2}$:
\begin{eqnarray}
\lefteqn{\mbox{Re}\Pi^{0}(q,\omega)=
\frac{1}{2\pi\sin\frac{qa}{2}\sqrt{(\omega/2\sin\frac{qa}{2})^2-1}}\cdot}
\nonumber\\& &
(\arctan(\frac{\cos(\frac{qa}{2}-k_F)}{\sqrt{(\omega/2\sin\frac{qa}{2})^2-1}}-
 \arctan(\frac{\cos(\frac{qa}{2}+k_F)}{\sqrt{(\omega/2\sin\frac{qa}{2})^2-1}})
\end{eqnarray}
For $\omega=2\sin\frac{qa}{2}$:
\begin{eqnarray}
\mbox{Re}\Pi^{0}(q,\omega)=
\frac{1}{2\pi\sin\frac{qa}{2}}
(\frac{1}{\cos(\frac{qa}{2}+k_F)}-
 \frac{1}{\cos(\frac{qa}{2}-k_F)})
\end{eqnarray}
For $0<\omega<2\sin\frac{qa}{2}$:
\begin{eqnarray}
\lefteqn{\mbox{Re}\Pi^{0}(q,\omega)=
\frac{1}{2\pi\sin\frac{qa}{2}\sqrt{1-(\omega/2\sin\frac{qa}{2})^2}}\cdot}
\nonumber\\& &
\ln |\frac{\cos(\frac{qa}{2}-k_F)-\sqrt{1-(\omega/2\sin\frac{qa}{2})^2}}
	  {\cos(\frac{qa}{2}-k_F)+\sqrt{1-(\omega/2\sin\frac{qa}{2})^2}}
\cdot\frac{\cos(\frac{qa}{2}+k_F)+\sqrt{1-(\omega/2\sin\frac{qa}{2})^2}}
	  {\cos(\frac{qa}{2}+k_F)+\sqrt{1-(\omega/2\sin\frac{qa}{2})^2}}|^2
\end{eqnarray}
The imaginary part is:
\begin{eqnarray}
\lefteqn{
\mbox{Im}\Pi^0(q,\omega)=\frac{-i}{2\sqrt{4\sin^2\frac{qa}{2}-\omega^2}}
\cdot}\nonumber\\
& &\cdot\theta(\cos p\cos\frac{qa}{2}+\sin p\sin\frac{qa}{2}-\cos k_F)
   \theta(\cos k_F-\cos p\cos\frac{qa}{2}+\sin p\sin\frac{qa}{2})\nonumber\\
& &\cdot\theta(2\sin\frac{qa}{2}-\omega)|_
   {\cos p=\pm\sqrt{1-(\omega/2\sin\frac{qa}{2})^2}}
\end{eqnarray}

In 2D, the Ewald sum $\Delta V(q_x,q_y)$ has been given by Glasser [9]. But
it was expedient to input a slightly ``screened'' form of $\Delta V(q_x,q_y)$.
This does not affect the physics but simplifies the numerical work:
\begin{eqnarray}
\Delta V(q_x,q_y)=\sum_{(m,n)\neq(0,0)}^{(\pm M,\pm M)}
\frac{e^2}{a\sqrt{m^2+n^2}}e^{i(mq_xa+nq_ya)-q_c\sqrt{m^2+n^2}}
\end{eqnarray}
having a minimum $\Delta V(\pi,\pi)=-1.024 e^2/a$ for $M=10$, $q_c=0.69$
(the numbers adopted in our numerical simulations.) Similarly, the lattice
$\Pi^{0}(q_x,q_y,\omega)$ (a generalized Watson integral) is replaced by the
isotropic 2D Lindhard polarization function
$\Pi^{0}(\sqrt{q_x^2+q_y^2},\omega)$
[10] evaluated in the effective mass approximation, and the filling factor
is determined by $n=k_F^2/4\pi$ for $n<\frac{1}{2}$ (electrons)
and $1-k_F^2/4\pi$ for $n>\frac{1}{2}$ (holes.) These approximations
should be good for all almost empty or almost full band, but should not be
considered accurate at or near half-filling.

Only two parameters govern our problem: one is the particle density
$n$. The other is $U/W=e^{2}/2ta$, measuring
the Coulomb interaction strength $U$ relative to half band width $W$.

The solution $\alpha$ is presented in Fig.1 for 1D case, and
Fig.2 for 2D case for $n<\frac{1}{2}$. The results are symmetric about
$n=\frac{1}{2}$, so $n>\frac{1}{2}$ is not shown.

As we see in Fig.1, solutions $\alpha$'s never reach $1.386$, the maximum
of $-\Delta V$ in 1D. This implies that causality can not be satisfied as
there always exist some $q_{c}<\pi/a$ such that for $q>q_{c}$, $V(q)<0$
for all density $n$ and interaction $U/W$. Thus we see that the 1D spinless
homogeneous Fermion gas is unstable. This instability is already known
by way of the exact solution by Mattis and Lieb [11] of the Luttinger
model. At arbitrarily small values of the interaction energy, these authors
found that the discontinuity of $<n_k>$ at the Fermi surface disappears,
and the Fermi liquid is transmuted into a ``Luttinger liquid'' with a
corresponding charge in correlation functions.

Fig.2 shows the situation in 2D; the $\alpha$'s increase along with $n$,
and finally exceed $1.024$, the maximum of $-\Delta V$ in our calculations,
after which they satisfy causality. Hence, at a given interaction strength
$U/W$, the spinless Fermion gas is {\em unstable only} at density $n$ lower
than some critical $n_c$. The phase diagram in Fig.3 shows $n_c$ v.s. $U/W$.
As $U/W$ increases, $n_{c}$ approaches $\frac{1}{2}$. An additional possible
region of instability against dimerization near $n=\frac{1}{2}$ can not be
investigated within the effective mass approximation. Thus it is possible
that at sufficiently large $U/W$, the 2D Fermi liquid becomes altogether
unstable, but due to the errors in obtaining the optimum $\alpha$ at large
$n$ using the effective mass approximation, and because of the systematic
and inherent errors in the RPA at large $U/W$, we can only speculate on the
strong-coupled behavior. Efros [12], who {\em has} studied the
strong-coupling limit of the 2D electron liquid (in a strong magnetic field
which effectively quenches the spin degree of freedom) has found
that in the absence of motional energy, the electron-electron interaction
leads to a very inhomogeneous phase. His result may be complementary to our
RPA approach, which was predicated on the weak-coupling regime.

The effects of introducing a local field correction [8] $G(q)$ were also
investigated, but were found to have no qualitative influence on our results.

In conclusion, we find for spinless Fermions that the Landau Fermi-Liquid
phase, characterized by a one-to-one correspondence with the Fermi gas,
including the discontinuity in $<n_k>$ at $k_F$, has only a restricted region
of validity in 2D, and that it is {\em never} applicable in 1D. This last
result was already anticipated in the solution of the 1D Luttinger liquid [11],
albeit from a different point of view. In the present paper, we reach these
conclusions by optimizing RPA using a Lagrange multiplier $\alpha$. At the
optimum $\alpha$, the RPA is found to satisfy the Pauli principle, thereby
becoming a simple and satisfactory tool with which to study the stability
of the Fermi liquid.

The phase diagrams for SU(2) Fermions in 2D and in 3D are more complex
and require a subtler analysis. They will be examined separately.

\newpage
\begin{center}
Figure Captions
\end{center}

FIG.1\\
\noindent{Solution $\alpha$ (in unit of $e^2/a$) required to satisfy
$g_{\alpha}(0)=0$ in 1D. Solid line is for $U/W=1.0$, dash line for $U/W=0.1$.
Causality requires $\alpha >1.386$, which can never be satisfied.}
\\

FIG.2\\
\noindent{Solution $\alpha$ (in unit of $e^2/a$) for $g_{\alpha}(0)=0$
in 2D. Solid line is for $U/W=1.0$, dash line for $U/W=0.1$. Causality
requires $\alpha >1.024$, as shown by horizontal dash line. It fails to be
satisfied at low density.}
\\

FIG.3\\
\noindent{Phase diagram in 2D, indicating maximum region of stability
of Fermi liquid (there may be an additional dimerization instability near
$n=\frac{1}{2}$ which we are unable to examine by our methods.)}
\\

\end{document}